\documentclass[12pt,english]{article}
\usepackage{geometry}
\usepackage{longtable}
\usepackage{booktabs}
\usepackage{multirow}
\usepackage{setspace}
\usepackage[authoryear,round,colon, sort&compress]{natbib}
\bibpunct{(}{)}{,}{autheryear}{,}{,}
\usepackage{rotating}

\usepackage[english]{babel}
\usepackage{amsmath, amsthm, amssymb, amscd, amsfonts, bm, bbm, mathrsfs}   % math packages
\usepackage{graphicx}
\usepackage{enumerate}
\usepackage[shortlabels]{enumitem}
\usepackage{url} % not crucial - just used below for the URL 
\usepackage[colorlinks]{hyperref}
\usepackage{xr-hyper}
\usepackage[capitalize,nameinlink,noabbrev]{cleveref} % to emulate \autoref style
\usepackage[algo2e,linesnumbered,lined,ruled]{algorithm2e}

\usepackage[bottom]{footmisc}
\usepackage[dvipsnames]{xcolor}
\hypersetup{
	breaklinks,
	linkcolor=blue,
	urlcolor=blue,
	anchorcolor=blue,
	citecolor=blue
}

\makeatletter

\theoremstyle{plain}

\@ifundefined{date}{}{\date{}}
\usepackage{babel}
\allowdisplaybreaks

\usepackage{tikz}
\usetikzlibrary{fit,positioning}

\allowdisplaybreaks

\newcommand*{\indep}{%
  \mathbin{%
    \mathpalette{\@indep}{}%
  }%
}
\newcommand*{\nindep}{%
  \mathbin{%                   % The final symbol is a binary math operator
    \mathpalette{\@indep}{\not}% \mathpalette helps for the adaptation
  }%
}
\newcommand*{\@indep}[2]{%
  \sbox0{$#1\perp\m@th$}%        box 0 contains \perp symbol
  \sbox2{$#1=$}%                 box 2 for the height of =
  \sbox4{$#1\vcenter{}$}%        box 4 for the height of the math axis
  \rlap{\copy0}%                 first \perp
  \dimen@=\dimexpr\ht2-\ht4-.2pt\relax
  \kern\dimen@
  {#2}%
  \kern\dimen@
  \copy0 %                       second \perp
}

\newtheorem{thm}{Theorem}

\theoremstyle{definition}
\newtheorem{rem}{Remark}

\newtheorem{eg}{Example}
\makeatletter
\newcommand*{\rom}[1]{\expandafter\@slowromancap\romannumeral #1@}
\newcommand{\HUGE}{\@setfontsize\Huge{40}{50}}   
\makeatother

\makeatletter
\newcommand{\labelcustom}[2]{%
	\protected@write \@auxout {}{\string \newlabel {#1}{{#2}{\thepage}{#2}{#1}{}} }%
	\hypertarget{#1}{#2}
}
\makeatother

\makeatletter
\newcommand{\labeltext}[3][]{%
	\@bsphack%
	\csname phantomsection\endcsname% in case hyperref is used
	\def\tst{#1}%
	\def\refmarkup{}%
	\ifx\tst\empty\def\@currentlabel{\refmarkup{#2}}{\label{#3}}%
	\else\def\@currentlabel{\refmarkup{#1}}{\label{#3}}\fi%
	\@esphack%
	\labelmarkup{#2}% visible printed text.
}
\makeatother

\makeatletter
\newcommand{\bianca}{\renewcommand\NAT@open{[}\renewcommand\NAT@close{]}}
\makeatother

\newcommand{\pr}{\mathsf{P}}
\newcommand{\eo}{\mathsf{E}}

\newcommand{\nd}{\mathsf{N}}

\newcommand{\ap}{\alpha} %alpha
\newcommand{\g}{\gamma} %gamma
\newcommand{\ga}{\Gamma} %Gamma
\newcommand{\dt}{\delta} % small delta
\newcommand{\Dt}{\Delta} % BIG Delta
\newcommand{\e}{\varepsilon} % epsilon
\newcommand{\s}{\sigma} % sigma
\newcommand{\Ld}{\Lambda} % BIG Lambda

\newcommand{\HH}{\mathbb{H}} % H
\newcommand{\R}{\mathbb{R}} % R
\newcommand{\dd}{\mathcal{D}}	% D
\newcommand{\ww}{\mathcal{W}}	% W
\let\oldnl\nl
\newcommand{\nlnonumber}{\renewcommand{\nl}{\let\nl\oldnl}}

\renewcommand{\eqref}[1]{(\ref{#1})}

\renewcommand{\sectionautorefname}{Section}
\renewcommand{\subsectionautorefname}{Section}
\renewcommand{\subsubsectionautorefname}{Section}
\renewcommand{\algorithmautorefname}{Algorithm}

\begin{document}

	\renewcommand{\sectionautorefname}{Section}
	\renewcommand{\subsectionautorefname}{Section}
	\renewcommand{\subsubsectionautorefname}{Section}
	\renewcommand{\algorithmautorefname}{Algorithm}

\title{
	Wild bootstrap for mean response inference \\
	in functional linear regression models
}
\author{
	Hyemin Yeon\thanks{Department of Mathematical Sciences, Kent State University},
	Xiongtao Dai\thanks{Division of Biostatistics, University of California, Berkeley},
	and Daniel John Nordman\thanks{Department of Statistics, Iowa State University} 
}

\newpage

\maketitle

%\begin{center}
%	\date{\today}
%\end{center}

\begin{abstract}
	
	\noindent
	Functional regressors  complicate inference in linear regression problems so that the bootstrap can play a useful role in quantifying uncertainty and calibrating intervals. 
	The best bootstrap  in practice, though, can depend on factors in the data as well as computational considerations and  
	existing bootstraps can have  limitations:
	residual bootstrap is computationally fast and simple but may fail when the errors are heterogeneous, 
	while paired bootstrap applies more generally in functional linear regression at a cost of much higher computation.   
	To bridge this gap, we develop a   wild bootstrap method for functional linear regression, 
	which is akin to a modified version of residual bootstrap but designed to have a wide scope of application like paired bootstrap, including to heteroscedastic errors. 
	Its theoretical consistency is established
	and  numerical studies suggest that wild bootstrap can provide accurate and computationally fast inference.
	Importantly, we also suggest a practical and effective approach of selecting truncation levels, specifically designed for mean response inference problems.
	The proposed bootstrap in functional linear regression  is further illustrated through a weather data example, and 
	an accompanying R package \texttt{BTSinFLRM}   provides numerical implementations.

\end{abstract}

%\newpage
%\tableofcontents

\newpage

% ================= Section 1 =================================

\section{Introduction} \label{WBsec1}

\subsection{Motivation}

Bootstrap methods have been widely used and studied for statistical inference in classical   linear regression models (LRMs) given as
\begin{align} \label{WBeq_MLRM}
	Y = \ap + \beta^\top X + \e.
\end{align}
where $Y$ and $X$ respectively represent a scalar response and a finite-dimensional vector regressor involving an
intercept $\ap$ and the slope vector $\beta$, while $\e$ denotes an error.
There are three main bootstrap methods for use in LRMs  (i.e., residual, paired, and wild bootstraps),
which all   aim to re-create  bootstrap versions $\{(Y_i^*,X_i^*)\}_{i=1}^n$ of observed data 
$\{(Y_i,X_i)\}_{i=1}^n$   for approximating sampling distributions.
\textit{Residual bootstrap}  (RB)  uses  the same regressors $X_i^* = X_i$ as the original data and generates a response as $Y_i^* = \widehat{\mu}(X_i) + \varepsilon_i^*$  
by using an estimator   $ \widehat{\mu}(X_i)$ in place of  
$\mu(X_i) \equiv \ap + \beta^\top X_i$ 
and  drawing a bootstrap error $\varepsilon_i^*$   from  centered residuals $\hat{\e}_i \equiv Y_i - \widehat{\mu}(X_i)$,
while  
{\it paired bootstrap (PB)}  creates each bootstrap observation $(Y_i^*,X_i^*)$   by drawing independently 
from the original data pairs $\{(Y_j,X_j)\}_{j=1}^n$  
\cite[cf.][]{Efron79, free81}. RB is 
computationally simple but strictly intended for  for homoscedastic regression models,
while PB  is valid   even under heteroscedastic 
models due to its 
nonparametric resampling, though 
the latter  often can entail  higher computational burdens, bias in bootstrap statistics, and potentially wider intervals compared to RB.     
{\it Wild bootstrap (WB)} was proposed  
to circumvent the respective limitations of RB and PB \cite[cf.][]{liu88, wu86}.
Like RB, WB uses the original data regressors $X_i^* = X_i$  and  generates a response as $Y_i^* = \widehat{\mu}(X_i) + \varepsilon_i^*$;
the key difference from RB is that 
the bootstrap error $\varepsilon_i^*$ in WB is defined in a way to depend  on $X_i$ so that WB can handle heteroscedasticity in regression.  
Unlike PB, though, WB has also easy implementation; 
for example, WB does not require computing an inverse covariance matrix of regressors in each bootstrap resample $\{(Y_i^*,X_i^*)\}_{i=1}^n$ as in PB.

Recent interest has   focused on extending  such bootstrap methods to inference about a functional linear regression model (FLRM) given by
\begin{align} \label{WBeq_FLRM}
	Y = \ap + \langle \beta, X \rangle + \e,
\end{align}
where $Y$ still denotes a scalar response (involving a model error)   but   $X$ is instead a  functional regressor.
That is, 
$X$ is a random element taking values in an infinite dimensional Hilbert space $\HH$ with inner product $\langle \cdot, \cdot \rangle$,
while the parameter $\beta \in \HH$ is now called the slope function  for determining the mean response $\mu(X)\equiv \ap + \langle \beta, X \rangle$.  
Unlike with a classical  LRM,  
the regressors $X$ in a FLRM can assume an infinite dimension of values, which creates challenges  in estimation, related to bias and scaling
\cite[cf.][]{CMS07, HH07}, and also complicates bootstrap inference.  Due to the latter, fewer formal bootstrap developments exist for FLRMs compared to standard LRMs,  
though some progress has been made.   
In FLRMs under homoscedasticity, RB  has been considered for interval estimation of the conditional mean, or similarly the projection $\langle \beta, X_0 \rangle$, of a new regressor $X_0$  by  \cite{GM11} and  \cite{YDN23RB}.  Taking a different perspective from mean response inference, 
\cite{KH24} % and \cite{LL25}   
proposed an RB approach for testing the significance of the slope $\beta$
under homoscedastiticy. 
Under  heteroscedastic  errors, 
\cite{YDN24PB} proposed a PB in FLRMs with  bias correction,
%\cite{YDN24PB} studied  PB in FLRMs, but found that  a PB version for FLRMs cannot directly follow the fashion of PB for MRLMs  (i.e., due to bias issues). Instead, 
%they proposed a modified PB to provide invalid inference in FLRMs, 
albeit with high computational costs.
%because the modified PB method requires solving new systems of estimating equations with each bootstrap resample involving functional regressors.
However, in contrast to LRMs,   formal inference with WB for even mean inference
has not established for FLRMs under possible heteroscedastic 
errors, which represents a methodological and theoretical gap for bootstrap inference under FLRMs.

% It is worth noting that, beyond confidence intervals, bootstrap methods can be applied to hypothesis tests as well (e.g.,~\cite{GGMG12, YDN24PB, KH24}); see \autoref{WBsec5} for more discussion on bootstrap for hypothesis tests in FLRMs.

\subsection{Our contributions} \label{WBssec_1_2}
This work aims to formally treat a wild bootstrap (WB) for inference for the mean response $\mu(X_0) \equiv \ap + \langle \beta, X_0 \rangle$ at a new regressor $X_0$ in FLRMs. Similar to LRMs,
the motivation for WB is that this approach has the potential to integrate the benefits of previous bootstrap methods for FLRMs.   
That is, WB can be valid for FLRMs under general assumptions (including heteroscedastic errors) much like PB, 
but WB also has potential to be computationally fast, like RB.  
Particularly with functional regressors, WB requires less computational  effort and is more computationally scalable  than PB.

	%We summarize our unique contributions.
	% (e.g., as WB does not require spectral decompositions for every bootstrap resample).
	To have a large scope of application, we prove the consistency of WB for mean response inference in FLRMs  under mild assumptions.
	Numerical  studies  between  our proposed WB and previous bootstraps (e.g., RB and PB) indicate that WB performs comparatively well.  WB similarly exhibits good coverage accuracy  under homoscedastic errors like RB  but, while RB completely fails to maintain coverage under heteroskedastic errors,  WB  also  has a coverage performance closer to PB 
	under heteroskedasticity at a lower computational cost.  Further, WB can show  
	much better coverage than PB in  cases where errors are both heterogeneous and highly skewed.  While all existing bootstrap methods for FLRMs require some tuning parameter selection (i.e., truncation levels), WB additionally allows an intuitive and practical method  for selecting these in practice.   The main idea is to select such truncation levels by examining where WB intervals stabilize in both center and length, which is not a feasible strategy for other bootstraps, like PB \citep{YDN24PB}, due to greater computational burdens and sensitives as  truncation levels vary.  This approach also allows tuning parameters to change with the  regressor $X_0$ at which an interval estimator of a target mean $\mu(X_0)$ is desired, which is not applicable in other classical choices for tuning parameters, such as  cross-validation \citep{HTF09} or fraction of variance explained \citep{KR17}.

	To highlight some distinctions from previous bootstrap works with FLRMs,  the development  of WB here under possible heteroscedasticity  involves a different theoretical treatment than with PB 
	(\autoref{WBthmWB}) due to the unique resampling scheme of WB.  For example, while the
	bootstrap observations from PB are exchangeable,
	such exchangeability does not hold for the WB data (e.g.,  $(Y_1^*, X_1) \overset{\mathsf{d}}{\neq}(Y_2^*, X_2)$) and    
	the WB responses $Y_i^*$ are defined in a distinct manner per  functional regressor $X_i$. 
	Consequently, theoretical validity must be approached differently (cf.~supplement~S1.3
%	\ref{WBssec_S_prelim_diffPB} 
	for details).
	Additionally, most previous contributions on resampling for FLRMs have again focused mainly on RB and homoscedastic errors, such as   
	\cite{GM11} and \cite{YDN23RB} for mean response  estimation and  
	\cite{KH24} and \cite{LL25} on slope testing (i.e., $\beta=0$).  We likewise target mean responses
	which enables mean comparisons 
	in a classically useful fashion under potential heteroscedasticity, as illustrated in \autoref{WBssec_4_2}, though the WB  could potentially be applied for  slope tests under heteroscedastic errors in future research.    
	Also, while \cite{GM11} considered a version of WB in addition to RB, our WB approach is quite different in that  (i) our general framework accommodates heteroscedasticity and the associated scaling/studentization required in FLRMs and (ii) we consider WB for inference about mean responses directly rather than incomplete or biased versions of means.
	In contrast to previous bootstraps, we also propose an intuitive data-driven method for selecting tuning parameters.

\subsection{Outline of the paper}

\autoref{WBsec2} 
reviews regression in FLRMs,   including  statistical quantities  involving mean parameters and projections.  
\autoref{WBsec_WB} then describes the WB
method for inference and formally establishes its validity.  	Empirical coverage performances are investigated through numerical studies in \autoref{WBsec3},
wherein   previous bootstrap methods, RB and PB, are also compared.
Through a data example in \autoref{WBsec4},
we illustrate
the implementation of WB, including   practical selection of tuning parameters involved in the WB procedure. 
\autoref{WBsec5} offers   concluding remarks.
The WB method  for FLRMs is implemented in the R package \texttt{BTSinFLRM} available online,\footnote{\href{https://github.com/luckyhm1928}{https://github.com/luckyhm1928}} 
while proofs and further technical details are deferred to the supplement.

\section{Estimation  and inference in  FLRMs} \label{WBsec2}

\autoref{WBssec_2_1} describes basic estimation in FLRMs, including the   functional principal component regression (FPCR) estimator of the slope, while
\autoref{WBssec_2_2}  briefly reviews   scaled statistical quantities for mean responses   under possible heteroscedasticity.

\subsection{Functional principal component regression estimator} \label{WBssec_2_1}

To describe a theoretical framework for estimation in FLRMs,
we assume that the slope function $\beta$ in   \eqref{WBeq_FLRM} lies in  an infinite-dimensional separable Hilbert space $\HH$ equipped with inner product $\langle \cdot, \cdot \rangle$ and the induced norm $\|\cdot\|$ defined as $\|x\| \equiv \sqrt{\langle x, x \rangle}$ for $x \in \HH$. 
The functional regressor $X$ is a random element  in  $\HH$ with a finite second moment $\eo[\|X\|^2]<\infty$,
while the response $Y$ and error $\e$ are both scalar-valued random variables with $\eo[\e|X] = 0$.
Here we do not assume zero means for both $X$ and $Y$,   as occurs in some theoretical developments \cite[cf.][]{CMS07, HH07, YDN23RB};
see \cite{CH06, YDN24PB} for related discussions.
%	The zero mean conditions for both $X$ and $Y$ can be relaxed by adding the intercept term, but are commonly assumed for ease of theoretical development \cite[cf.][]{CMS07, HH07, YDN23RB};
%	see  \cite{CH06, YDN24PB} for related discussions.

Similar to how a least squares estimator in a classical LRM \eqref{WBeq_MLRM} is defined by normal equations,
a functional version of   normal equations   is also   important   to  defining estimation   in a FLRM \eqref{WBeq_FLRM}.
Write $\ga \equiv \eo[(X-\eo[X]) \otimes (X-\eo[X])]$ for the covariance operator of $X$, 
where $x \otimes y:\HH\to\HH$ denotes the tensor product between two elements $x,y \in \HH$
defined as $(x\otimes y)(z)=\langle z, x \rangle y \in \HH$ for $z \in \HH$.
Denoting $\Dt \equiv \eo[(Y-\eo[Y])(X-\eo[X])]$ as the cross-covariance function between $X$ and $Y$, we obtain   normal equations from \eqref{WBeq_FLRM} as 
$\Dt = \ga \beta$.
The slope function is then identifiable, and hence given as $\beta = \ga^{-1} \Dt$, under  a condition $\{ x \in \HH: \ga x = 0 \} = \{0\}$ \cite[cf.][]{CFS03, CMS07}.
%
%\begin{enumerate}[(A0)]
%	\item $\ker \ga = \{0\}$, where $\ker \ga \equiv . \label{condModelIdentif}
%\end{enumerate}  
%	Here, the inverse covariance operator $\ga^{-1}$ may by unbounded, but its image$

With a finite second moment assumption $\eo[\|X\|^2]<\infty$,
because the covariance operator $\ga \equiv \eo[(X-\eo[X]) \otimes (X-\eo[X])]$ is self-adjoint, non-negative definite, and compact \cite[Chapter~4]{HE15},
this admits a spectral decomposition as
$\ga  = \sum_{j=1}^\infty \g_j (\phi_j \otimes \phi_j)$,
where $\g_j$ and $\phi_j$ respectively denote the $j$-th eigenvalue and eigenfunction of $\ga$ for an integer $j \geq 1$ \cite[Theorem~7.2.6]{HE15}.
The set $\{\phi_j\}_{j=1}^\infty$ of eigenfunctions forms an orthonormal system of $\HH$,
and the eigenvalues $\{\g_j\}_{j=1}^\infty$ are a non-negative, non-increasing sequence such that $\g_j \to 0$ as $j \to \infty$.
Writing  $\ga^{-1} \equiv \sum_{j=1}^\infty \g_j^{-1} (\phi_j \otimes \phi_j)$, the slope function in \eqref{WBeq_FLRM}  can be expressed as
\begin{align} \label{WBeq_beta}
	\beta = \ga^{-1} \Dt = \sum_{j=1}^\infty \g_j^{-1} \langle \Dt, \phi_j \rangle \phi_j.
\end{align}

Based on a random sample $\{(Y_i, X_i)\}_{i=1}^n$ with $Y_i = \ap + \langle \beta, X_i \rangle +\e_i$  from the model \eqref{WBeq_FLRM}, 
the so-called FPCR estimator  of  $\beta$ can then be stated as a truncated sample version of the slope function \eqref{WBeq_beta}.
To explain, define the sample counterparts of $\ga$ and $\Dt$ as $\hat{\ga}_n \equiv n^{-1} \sum_{i=1}^n (X_i - \bar{X})^{\otimes 2}$ and $\hat{\Dt}_n \equiv n^{-1} \sum_{i=1}^n (Y_i - \bar{Y}) (X_i - \bar{X})$, 
where $\bar{X} \equiv n^{-1} \sum_{i=1}^n X_i$ and $\bar{Y} \equiv n^{-1} \sum_{i=1}^n Y_i$ denote  the sample means of $\{X_i\}_{i=1}^n$ and $\{Y_i\}_{i=1}^n$, and $x^{\otimes 2} \equiv x \otimes x$ for $x \in \HH$.
The sample covariance operator $\hat{\ga}_n$ also admits spectral decomposition as $\hat{\ga}_n \equiv \sum_{j=1}^n \hat{\g}_j \hat{\phi}_j^{\otimes 2}$, 
where $\hat{\g}_j$ and $\hat{\phi}_j$ are the $j$-th sample eigenvalue and sample eigenfunction.
However, in contrast to inversion of $\ga$ as in \eqref{WBeq_beta},
inversion of  $\hat{\ga}_n$ is ill-posed 
because $\hat{\ga}_n$ has finite rank in the sense that the image of $\hat{\ga}_n$ has finite dimension.
This technical complication can be overcome by regularizing the inversion of $\hat{\ga}_n$ 
and obtaining the FPCR estimator $\hat{\beta}_{h_n}$ of $\beta$   as
\begin{align} \label{WBeq_FPCRest}
	\hat{\beta}_{h_n} \equiv \hat{\ga}_{h_n}^{-1} \hat{\Dt}_n = \sum_{j=1}^{h_n} \hat{\g}_j^{-1} \langle \hat{\Dt}_n, \hat{\phi}_j \rangle \hat{\phi}_j,
\end{align}
where  $\hat{\ga}_{h_n}^{-1} \equiv \sum_{j=1}^{h_n} \hat{\g}_j^{-1} \hat{\phi}_j^{\otimes 2}$ is a finite approximation of $\ga^{-1} \equiv \sum_{j=1}^\infty \g_j^{-1} \phi_j^{\otimes 2}$,
and the truncation level $h_n$ represents the number of eigenpairs used in estimation \citep{CFS99, CH06, HH07, CMS07}.
The intercept parameter $\ap$ is then estimated by $\hat{\ap}_{h_n} \equiv \bar{Y} - \langle \hat{\beta}_{h_n}, \bar{X} \rangle$.

\subsection{Target statistical quantities  for mean response} \label{WBssec_2_2}

While the FPCR estimator $\hat{\beta}_{h_n}$ is consistent \cite[cf.][Theorem~S3]{YDN24PBsupp},  
seminal work by \cite{CMS07} has shown that it is impossible for $a_n (\hat{\beta}_{h_n} - \beta)$ to converge in distribution to a non-degenerate random element taking values in $\HH$ for any diverging scaling sequence $\{a_n\}$.  
This feature complicates inference about the slope function $\beta$ directly.  
Importantly, however,  
meaningful inference about the conditional mean $\mu(X) \equiv \ap + \langle \beta, X \rangle$ 
of a response in the FLRM \eqref{WBeq_FLRM}
is possible by combining an estimator       
with an appropriate scaling term to define a statistical quantity with a well-defined distribution in large samples.
We next briefly review such target quantities for inference \citep{CMS07, YDN23RB, YDN24PB}. 
Let $X_0$ denote a new  regressor that is independent of the sample $\{(Y_i, X_i)\}_{i=1}^n$ and identically distributed as $X$. 
Then, a main problem of  interest is the inference about the true mean  
\begin{align}
	\mu(X_0) \equiv \ap + \langle \beta, X_0 \rangle \label{eqMeanTrue}
\end{align}
of a response given $X_0$,
which is estimated by its sample counterpart
\begin{align}
	\hat{\mu}_{h_n}(X_0) \equiv \hat{\ap}_{h_n} + \langle \hat{\beta}_{h_n}, X_0 \rangle. \label{eqMeanEst}
\end{align}
We can  also  consider related   inference about the projection $\langle \beta, X_0 - \eo[X] \rangle$ of the slope function $\beta$ onto a new predictor $X_0 - \eo[X]$ centered by the regressor mean $\eo[X]$.  This projection is useful for 
assessing the  effect  $\langle \beta, X_0 \rangle$ of the slope at a new predictor $X_0$ relative to an overall slope effect $\langle \beta, \eo[X] \rangle$, in which estimation of the  model intercept in (\ref{eqMeanTrue})  (or the  unconditional mean response  $\eo[Y]$ in $\alpha$) also becomes unnecessary.  We then consider statistical quantities based on differences between estimated and true mean responses or projections given by 
\begin{eqnarray} \label{WBeq_Tn}
	T_{\mathrm{mean},n}(X_0) 
	&\equiv&  
	\sqrt{n \over \hat{s}_{h_n}(X_0)} \{  \hat{\mu}_{h_n}(X_0) - \mu(X_0) \},\\ \nonumber
	T_{\mathrm{proj},n}(X_0) &\equiv&  
	\sqrt{n \over \hat{s}_{h_n}(X_0)} (\langle \hat{\beta}_{h_n}, X_0 - \bar{X} \rangle - \langle \beta, X_0 - \eo[X] \rangle) 
\end{eqnarray}
where   $\mu(X_0)$  and $\hat{\mu}_{h_n}(X_0)$  are from  \eqref{eqMeanTrue}-\eqref{eqMeanEst}.
%	or for the difference between estimated and true (centered) projections
%	\begin{align} \label{WBeq_Tn_proj}
	%		T_{\mathrm{proj},n}(X_0) 
	%		\equiv \sqrt{n \over \hat{s}_{h_n}(X_0)} (\langle \hat{\beta}_{h_n}, X_0 - \bar{X} \rangle - \langle \beta, X_0 - \eo[X] \rangle).
	%	\end{align}
Here $\sqrt{n/\hat{s}_{h_n}(X_0)}$ denotes an appropriate
scaling factor with general  and possibly heteroscedastic FLRMs, as defined by  
\begin{align} \label{WBscalingHeteroHat}
	\hat{s}_{h_n}(x) \equiv \langle \hat{\Ld}_n \hat{\ga}_{h_n}^{-1} (x-\bar{X}), \hat{\ga}_{h_n}^{-1} (x-\bar{X}) \rangle = \|\hat{\Ld}^{1/2} \hat{\ga}_{h_n}^{-1} (x-\bar{X}) \|^2, \quad x \in \HH,
\end{align}
in terms of 
a sample covariance operator
\begin{align*}
	\hat{\Ld}_n \equiv n^{-1} \sum_{i=1}^n \left( X_i \hat{\e}_{i,k_n} - n^{-1} \sum_{i=1}^n X_i \hat{\e}_{i,k_n} \right)^{\otimes 2}
\end{align*} 
based on 
residuals 
\begin{align} \label{WBeq_residuals}
	\hat{\e}_{i,k_n} \equiv Y_i - \hat{\mu}_{k_n}(X_i), \quad i=1,\dots,n
\end{align}
with $\hat{\mu}_{k_n}(X_i) \equiv \hat{\ap}_{k_n} + \langle \hat{\beta}_{k_n}, X_i \rangle$
formed by a FPCR estimator $\hat{\beta}_{k_n}$ in \eqref{WBeq_FPCRest} using a truncation level $k_n$;
for greatest generality here, the truncation level $k_n$ used for defining residuals  need not be the same as the truncation $h_n$
for the target FPCR estimator $\hat{\beta}_{h_n}$   
and the sample covariance operator  
$\hat{\Ld}_n$ will be consistent 
for $\Ld$ whenever $\hat{\beta}_{k_n}$ is consistent for $\beta$ \cite[cf.][Lemma~S23]{YDN24PBsupp}. 
The scaling $\hat{s}_{h_n}(x)$ in \eqref{WBscalingHeteroHat} estimates its population counterpart
\begin{align} \label{WBscalingHetero}
	s_{h_n}(x) 
	\equiv \langle \Ld \ga_{h_n}^{-1} (x-\eo[X])  \ga_{h_n}^{-1} (x-\eo[X]) \rangle 
	= \|\Ld^{1/2} \ga_{h_n}^{-1} (x-\eo[X]) \|^2, \quad x \in \HH, 
\end{align}
with $\Ld \equiv \eo[(X\e)^{\otimes 2}]$ \cite[cf.][Proposition~S4]{YDN24PBsupp},
where this scaling for FLRMs can be motivated by recalling  that a  standard normal limit follows in a classical  
LRM \eqref{WBeq_MLRM} for a scaled LRM projection 
$\sqrt{n/s(X_0)} \{(\hat{\beta}^{LSE})^\top X_0 - \beta^\top X_0\}$, involving the least squares estimator  $\hat{\beta}^{LSE}$ of the slope $\beta \in \R^p$, a new predictor $X_0\in \R^p$,  and  analog scaling as 
$s(x) \equiv x^\top \ga^{-1} \Ld \ga^{-1} x =  \|\Ld^{1/2}\ga^{-1}x\|_{\R^p}^2$, $x \in \R^p$,
with $\ga \equiv \eo[XX^\top]$ and $\Ld \equiv \eo[(X\e)(X\e)^\top]$ \citep{free81}.  
For the FLRMs, 
the inverse covariance operator $\ga^{-1}$ is first truncated due to its unboundedness, 
which  leads  to analog scaling $s_{h_n}(x)$ for the heteroscedastic FLRMs as in \eqref{WBscalingHetero}.

%	The scaling terms in \eqref{WBscalingHeteroHat}-\eqref{WBscalingHetero} can be reduced 
%	if the errors have constant variance as $\eo[\e^2|X] \equiv \s_\e^2 \in (0,\infty)$.
%	In this case, we have $\Ld = \s_\e^2 \ga$ so that $s_{h_n}(X_0) = \s_\e^2t_{h_n}(X_0)$ holds with respect to
%	the scaling $t_{h_n}(X_0) \equiv \|\ga_{h_n}^{-1/2}(X_0-\eo[X])\|^2$ for homoscedastic FLRMs; 
%	additionally, we estimate it with  $\hat{\s}_\e^2 \hat{t}_{h_n}(X_0)$, 
%	based on the sample counterpart $\hat{t}_{h_n}(X_0) \equiv \|\hat{\ga}_{h_n}^{-1/2}(X_0-\bar{X})\|^2$ of $t_{h_n}(X_0)$
%	and the average $\hat{\s}_\e^2 \equiv n^{-1} \sum_{i=1}^n \hat{\e}_{i,k_n}^2$ of the squared residuals \eqref{WBeq_residuals}.
%	Here, $\hat{\ga}_{h_n}^{-1/2} \equiv \sum_{j=1}^{h_n} \hat{\g}_j^{-1/2} (\hat{\phi}_j \otimes \hat{\phi}_j), \ga_{h_n}^{-1/2} \equiv \sum_{j=1}^{h_n} \g_j^{-1/2} (\phi_j \otimes \phi_j)$ denote square-root operators of $\hat{\ga}_{h_n}^{-1} \equiv \sum_{j=1}^{h_n} \hat{\g}_j^{-1} (\hat{\phi}_j \otimes \hat{\phi}_j), \ga_{h_n}^{-1} \equiv \sum_{j=1}^{h_n} \g_j^{-1} (\phi_j \otimes \phi_j)$, respectively.
Central limit theorems (CLTs) in homoscedastic FLRMs have been  studied by \cite{CMS07} and \cite{YDN23RB}, who work with the projection statistic $T_{\mathrm{proj},n}(X_0)$ in \eqref{WBeq_Tn} upon  replacing $\hat{s}_{h_n}(X_0)$ from \eqref{WBscalingHeteroHat} by a homoscedastic version. However, sampling distributions and the scaling form of projection statistics 
can change under 
under 
heteroscedastic FLRMs in general.  For this reason, we follow
the framework of 
\cite{YDN24PB}  for defining   the scaled quantities in \eqref{WBeq_Tn}, which  admit large-sample standard normal limits for either homoscedastic or hetereroscedastic FLRMs.
For completeness, we provide this  CLT in Theorem~S1 of the supplement.

\section{Wild bootstrap (WB) in heteroscedastic FLRMs} \label{WBsec_WB}

%\autoref{WBssec_2_3} describes a  WB approach for inference in FLRMs under potential hetereroscedasticity, 
%while \autoref{WBssec_2_4} establishes the method's theoretical validity.  
%
%
%\subsection{WB inference for mean response} \label{WBssec_2_3}

From the Introduction, recall the motivation for WB with FLRMs is that this approach can find a compromise of strengths in existing bootstrap approaches, i.e., RB is computationally fast but  valid   only for homoscedastic FLRMs, while  PB  is more generally valid but can be   conservative   (wider intervals) and computationally expensive for large data  due to its fully non-parametric resampling.     In contrast to RB and PB, 
the WB aims re-creates a bootstrap sample in which  regressors match those $\{X_i\}_{i=1}^n$ of the original data (i.e., like RB which is beneficial for computation) {\it and} 
bootstrap responses $\{Y_i^*\}_{i=1}^n$ are formed by  resampling residuals in a way that allows for heteroscedastic behavior (i.e., which is advantageous for wide validity).  Namely, the WB method creates a data sample $\dd_n^* \equiv \{(Y_i^*,X_i)\}_{i=1}^n$ of regressor-response pairs, in which bootstrap responses are  generated as  
\begin{align*}
	Y_i^* = \hat{\mu}_{g_n}(X_i) + \e_i^*, \quad i=1, \dots, n
\end{align*}
with $\hat{\mu}_{g_n}(X_i) \equiv \hat{\ap}_{g_n} + \langle \hat{\beta}_{g_n}, X_i \rangle$
formed by a FPCR estimator $\hat{\beta}_{g_n}$ in \eqref{WBeq_FPCRest} using a truncation level $g_n$;   for greatest generality, the truncation level $g_n$ for defining the WB version $\hat{\mu}_{g_n}(X_i)$ of the response mean $\mu(X_i)$  need not be the 
same   as the truncation $h_n$
defining the target FPCR estimator $\hat{\beta}_{h_n}$ 
from  (\ref{WBeq_FPCRest})  or the truncation $k_n$ 
defining residuals $\{\hat{\e}_{i,k_n}\}_{i=1}^n$ from (\ref{WBeq_residuals}).
WB errors $\{\e_i^*\}_{i=1}^n$ are generated independently, though not identically as with RB, 
in a way that these have mean zero  $\eo^*[\e_i^*] = 0$ and a variance   $\eo^*[(\e_i^*)^2] = \hat{\e}_{i,k_n}^2$ in the bootstrap world,
where  $\eo^*$ denotes expectation under the bootstrap generation and  $\hat{\e}_{i,k_n}$ denotes 
the $i$-th (original data) residual
defined in \eqref{WBeq_residuals}, $i=1,\ldots,n$.
Some specifications for WB errors $\{\e_i^*\}_{i=1}^n$  are given below.

From these bootstrap data, the WB analog of the FPCR estimator   $\hat{\beta}_{h_n}\equiv \hat{\ga}_{h_n}^{-1} \hat{\Dt}_n$  from \eqref{WBeq_FPCRest} is given as 
$\hat{\beta}_{h_n}^* \equiv \hat{\ga}_{h_n}^{-1} \hat{\Dt}_n^*$,
where $\hat{\Dt}_n^* \equiv n^{-1} \sum_{i=1}^n (Y_i^* - \bar{Y}^*)(X_i - \bar{X})$ denotes the WB cross-covariance function between $\{Y_i^*\}_{i=1}^n$ and $\{X_i\}_{i=1}^n$ with $\bar{X} \equiv n^{-1} \sum_{i=1}^n X_i$ and $\bar{Y}^* \equiv n^{-1} \sum_{i=1}^n Y_i^*$,
while the WB intercept is similarly computed as $\hat{\ap}^* \equiv \bar{Y}^* - \langle \hat{\beta}_{h_n}^*, \bar{X} \rangle$;
we then construct the WB estimator of the conditional mean as
\begin{align} \label{eqMeanWB}
	\hat{\mu}_{h_n}^* (X_0) \equiv \hat{\ap}^* + \langle \hat{\beta}_{h_n}^*, X_0 \rangle.
\end{align} 
The WB then approximates the sampling distribution of 
a target statistical quantity $T_{\mathrm{mean},n}(X_0)$
or $T_{\mathrm{proj},n}(X_0)$ from \eqref{WBeq_Tn}
at a new predictor $X_0$   with a  bootstrap counterpart
\begin{eqnarray} \label{WBeq_TnStar}
	T_{\mathrm{mean},n}^*(X_0) 
	&\equiv&  
	\sqrt{n \over \hat{s}_{h_n}^*(X_0)} 
	\{\hat{\mu}_{h_n}^*(X_0) - \hat{\mu}_{g_n}(X_0)\},\\ \nonumber
	T_{\mathrm{proj},n}^*(X_0) &\equiv&  
	\sqrt{n \over \hat{s}_{h_n}^*(X_0)} 
	(\langle \hat{\beta}_{h_n}^*, X_0 - \bar{X} \rangle - \langle \hat{\beta}_{g_n}, X_0 - \bar{X} \rangle),
\end{eqnarray}
which involve 
scaling  $\hat{s}_{h_n}^*(X_0)$ computed from the bootstrap sample; regarding the latter,  
a collection of WB residuals is computed as 
\begin{align*} % \label{WBeq_WBresiduals}
	\hat{\e}_{i,k_n}^* \equiv Y_i^* - \hat{\mu}_{k_n}^*(X_i), \quad i=1,\dots,n,
\end{align*}
using the WB regression estimator $\hat{\beta}_{k_n}^* = \hat{\ga}_{h_n}^{-1} \hat{\Dt}_n^*$ with truncation level $k_n$, and then
\begin{align} \label{WBeq_WBsHatStar}
	\hat{s}_{h_n}^*(x) \equiv \langle \hat{\Ld}_n^* \hat{\ga}_{h_n}^{-1} (x-\bar{X}), \hat{\ga}_{h_n}^{-1} (x-\bar{X}) \rangle = \|(\hat{\Ld}_n^*)^{1/2} \hat{\ga}_{h_n}^{-1} (x-\bar{X}) \|^2, \quad x \in \HH,
\end{align}
defines the bootstrap scaling with
\begin{align*}
	\hat{\Ld}_n^* \equiv n^{-1} \sum_{i=1}^n \left( X_i \hat{\e}_{i,k_n}^* - n^{-1} \sum_{i=1}^n X_i \hat{\e}_{i,k_n}^* \right)^{\otimes 2}
\end{align*}
as an empirical covariance from the WB sample products $\{X_i \hat{\e}_{i,k_n}^*\}_{i=1}^n$.

In terms of constructing the WB errors $\{\e_i^*\}_{i=1}^n$, one popular approach is to use   random multipliers, which corresponds to the so-called \emph{multiplier WB}.    
Independently of the original data, let  $\ww_n \equiv \{W_i\}_{i=1}^n$ denote a set of independent random variables, generated for use in bootstrap to have   mean  zero and   variance   one.  
Then,   WB errors $\{\e_i^*\}_{i=1}^n$ are defined as 
$\e_i^* \equiv W_i \hat{\e}_{i,k_n}$, $i=1,\ldots,n$,
with $\hat{\e}_{i,k_n}$ denoting
the $i$-th (original data) residual
from \eqref{WBeq_residuals}.  By this construction, the WB errors possess the required moment properties in bootstrap:    $\eo^*[\e_i^*] = 0$ and    $\eo^*[(\e_i^*)^2] = \hat{\e}_{i,k_n}^2$.    
Our numerical studies show that the choice of   multipliers may  affect the performance of   WB, though the common choices of multipliers below tend to induce similar results, where differences  diminish with increasing sample size.  

\begin{eg} \label{egW}
	Basic choices of multipliers include a
	two-point distribution with $\pr(W_i = -(\sqrt{5}-1)/2) = (\sqrt{5}+1) / (2\sqrt{5}) = 1-\pr(W_i = (\sqrt{5}+1)/2)$ \citep{cao91, HM91marron, mam00}; a standard normal distribution with $W_i \sim \nd(0,1)$; and a mean zero distribution with second and third moments of one, e.g., $W_i = V_i/2+(V_i^2-1)/2$ for a variable $V_i \sim \nd(0,1)$ \citep{HM91, mam93}.
\end{eg}

%\subsection{Consistency of WB} \label{WBssec_2_4}

We now present a consistency result to 
%Under   mild technical  conditions,   
%\autoref{WBthmWB} 
verify that the WB method
can validly approximate the distribution of   target statistics  $T_{\mathrm{mean},n}(X_0)$ and $T_{\mathrm{proj},n}(X_0)$, respectively, for inference about the mean response $\mu(X_0)$ and projection $\langle \beta, X_0 - \eo[X] \rangle$.
Let $\pr^*$  denote bootstrap probability, as induced by  WB resampling  
given the data $\{(Y_i, X_i)\}_{i=1}^n$.

\begin{thm} \label{WBthmWB}
	In addition to the assumptions of the CLT in Theorem~S1 of the supplement, 
		we suppose that Conditions~(L), (R), and (W1)-(W2)
		(including Condition~(A4) for $g_n$) hold,
		which are listed in Section~S1.2 of the supplement.
	%	Suppose that Conditions~\ref{condMomentX}-\ref{condBasicTrunc}, \ref{condScalingHetero}, \ref{condMomentXe}, \ref{condBalanceGaLd}, \ref{condRatioTunes}, \ref{condLdHeterBTS} (including Condition~\ref{condBasicTrunc} for $k_n,g_n$) hold 
	%	%		Suppose  assumptions in \autoref{WBthmPB} hold
	%	along with Conditions~\ref{condWBmultiplier1}-\ref{condWBmultiplier2} and Condition~\ref{condBias} for some $u>7$,
	%	which are all listed in \autoref{WBssec_S_prelim_Cond} of the supplement.
	%	We further suppose that 
	%	$n^{-1/2} h_n^{7/2} (\log h_n)^4 = o(1)$
	%	and $n=O(m(h_n,u))$ hold.
	Then,  letting $T_n(X_0)$ denote either $T_{\mathrm{mean},n}(X_0)$ or $T_{\mathrm{proj},n}(X_0)$ from \eqref{WBeq_Tn} and $T_n^*(X_0)$ denote the WB counterpart from    \eqref{WBeq_TnStar}, the WB approximation of  $T_n(X_0)$ by  $T_n^*(X_0)$ is valid: as $n\to\infty$, 
	\begin{align*}
		\sup_{y \in \R} |\pr^*(T_n^*(X_0) \leq y| X_0) - \pr(T_n(X_0) \leq y| X_0)|  \xrightarrow{\pr} 0.
	\end{align*}
\end{thm}

Similar to    RB  results in \cite{YDN23RB}, 
the consistency of WB here could also be extended to cases where 
when the new predictor $X_0$ depends on the observed regressors $\{X_i\}_{i=1}^n$ 
or when the new $X_0$ and observed $\{X_i\}_{i=1}^n$ regressors do not share the same distribution, though we do not pursue repeating technical details here.  As an example, with assumptions/arguments as in \citet[Propositions~S3, S4, S6, and S7]{YDN23RB},
	\autoref{WBthmWB} holds for $X_0$   given as an average of some observed regressors $X_1, \dots, X_L$ for some $L \leq n$;
	see also \citet[Remark~7]{YDN24PB}.
	We have also relegated the
	technical conditions for \autoref{WBthmWB}
	to the supplement, with 
	full details  provided in Section~S1.2.
	Conditions~(W1)-(W2) above are
	WB error assumptions that hold  for the multipliers in \autoref{egW}.
	For illustration, we may provide a concrete example that satisfies the remaining process conditions.    
	In essence, these involve the eigenvalue decay of regressors, the smoothness of the slope function, the heteroscedastic error structure, and the growth rate for the truncation parameter $h_n$.

	\begin{eg} \label{egCond}
		%	To appreciate the technical assumptions, we provide a concrete example that satisfies all these conditions.
		In the space $\HH=L^2([0,1])$ of all square integrable functions from $[0,1]$ to $\R$,
		we consider the Fourier basis functions $\{\phi_j\}_{j=1}^\infty$ as eigenfunctions,
		while the corresponding eigenvalues $\{\g_j\}_{j=1}^\infty$
		are determined by their gaps $\dt_j \equiv \g_j-\g_{j+1} \asymp j^{-a}$ with decay rate $a > 2$ and $\g_1 \equiv \sum_{j=1}^\infty \dt_j$.
		The distribution of the random function $X$ is then defined through the (normalized) functional principal component (FPC) scores $\g_j^{-1/2} \langle X, \phi_j \rangle = \xi W_j$, 
		where $\{W_j\}_{j=1}^\infty$ are iid standard normal variables independent of the  latent random variable $\xi$ with $\eo[\xi^{10}]<\infty$. 
		This construction leads to dependence in principal component scores and satisfies Conditions~(A1)-(A4) along with Condition~(S). %\ref{condScalingHetero}.
		We next consider heteroscedastic errors with conditional variance $\s^2(X) \equiv \eo[\e^2|X] = \|X\|^2$,
		which fulfill Conditions~(C), (D), and (L);
%		\ref{condMomentXe}, \ref{condBalanceGaLd}, and \ref{condLdHeterBTS}; 
		for example, the log normal error as in the simulation studies (\autoref{WBsec3}) satisfies this.  
		A polynomial decay rate $b$ for the slope function $\beta$,
		i.e., $|\langle \beta, \phi_j \rangle| \asymp j^{-b}$,	together with truncation choice such that $h_n \asymp n^{1/v}$ with $\min\{7,2a+1\} < v < a+2b-1$ yields Condition~$B(u)$
%		 Condition~\ref{condBias}   
		with $u>7$ and the growth rate $n^{-1/2}h_n^{7/2} (\log h_n)^4 = o(1)$ for $h_n$ in \autoref{WBthmWB}. 
		%The bootstrap bandwidth Condition~\ref{condRatioTunes} also holds by taking $\tau \geq 1$ such as $h_n \asymp 2g_n$. 
		
	\end{eg}

% ================= Section 3 =================================
\section{Numerical studies} \label{WBsec3}

Here we   examine the performance of  confidence intervals constructed by WB  as well as by other bootstrap methods (i.e., RB and PB); these are implemented in the R package \texttt{BTSinFLRM} accompanying the paper.
In   numerical studies, we focus on inferring the projection $\mu(X_0) \equiv \langle \beta, X_0 \rangle$
by assuming both response and regressor have zero mean, for simplicity.   \autoref{WBssec_3_1}  describes the simulation design,
while  \autoref{WBssec_3_2} presents coverage rates and computational timings. 
Take-aways 
%and recommendations  between different bootstrap options 
are   summarized in \autoref{WBssec_3_3}.

\subsection{Simulation designs} \label{WBssec_3_1}

We  consider different scenarios to investigate the performance of  bootstrap methods.  
The slope function $\beta = \sum_{j=1}^J \beta_j \phi_j$ under consideration is defined by its Fourier series truncated up to a large integer $J=15$,
where $\{\beta_j\}$ denote the Fourier coefficients of $\beta$ and $\{\phi_j\}_{j=1}^J$ represents the first $J$ Fourier basis functions $\{1, \sin(2\pi t), \cos(2\pi t), \dots, \}$ on $[0,1]$.
Here, we define $\beta_j = 3W_{\beta, j}j^{-b}$ with $b=3.5$, 
where $\{W_{\beta, j}\}$ are independent variables with the identical distribution $\pr(W_{\beta, j}=-1) = 1/2 = \pr(W_{\beta, j}=1)$.	
The functional regressors $\{X_i\}_{i=1}^n$ are generated as independent copies of the random element $X$, which is constructed based on the following truncated Karhunen--Loève expansion
\begin{align} \label{WBeq_KLtrunc}
	X \overset{\mathsf{d}}{=} \sum_{j=1}^J \sqrt{\g_j} \xi_j \phi_j.
\end{align}
We take the Fourier basis functions $\{\phi_j\}_{j=1}^J$ for the eigenfunctions, while the eigenvalues are constructed based on the eigengaps with polynomial decay rate as 
$\dt_j = \g_j - \g_{j+1} = 2j^{-a}$ where $\g_1 = 2\sum_{j=1}^\infty j^{-a}$ and $a=2.5$. 
All functions above are evaluated at the same 50 equi-distant discretized points in $[0,1]$.  
For brevity, we consider only one combination of $a$ and $b$ as  above,
as these values are less consequential than the distribution of 
FPC scores and errors in the model  \citep{YDN23RB, YDN24PB}.  
%   The previous work by \cite{YDN23RB, YDN24PB} indicate that the bootstrap intervals work better with larger $a$ but its performance does not vary across different $b$.
As in Example~\ref{egCond}, the FPC scores $\{\xi_j\}_{j=1}^J$ are defined as
$\xi_j = \xi W_j$, where $\xi$ is a common latent and $\{W_j\}$ are independent standard normal random variables, which are independent of $\xi$;
this gives non-normal and dependent FPC scores $\{\xi_j\}_{j=1}^J$.
As for the distribution of the  latent variable $\xi$ that determines the moment of the regressor $X$, we focus  on a
$\mathsf{t}(4)$-distribution for  $\xi$; this represents a difficult case in the sense that  coverage accuracy of bootstrap intervals is expected to increase as $\xi$ has  higher moments (cf.~\citealp{YDN23RB, YDN24PB}).
The functional regressors $\{X_i\}_{i=1}^n$ and a new regressor $X_0$ are then generated following the expansion in \eqref{WBeq_KLtrunc}.

The data pairs $\{(Y_i, X_i)\}_{i=1}^n$ are generated with different sample sizes $n \in \{50, 200, 1000\}$, 
where the scalar responses $\{Y_i\}_{i=1}^n$ defined by error terms  $\e_i$ that follow a centered log normal distribution conditional on $X_i$.
The latter conditional distribution, written terms of a generic error $\e$ and regressor  $X$, is given by 
$\log (\e + \exp [\mu(X)+\tau^2/2]) \sim \nd(\mu(X), \tau^2)$, 
where the mean $\mu(X) \equiv \{\log [\s^2(X) / (e^{\tau^2}-1)] - \tau^2\}/2$
is defined as either $\s^2(X) = \mathrm{tr}(\ga) = \sum_{j=1}^J \g_j$ in the homoscedastic scenario or as $\s^2(X) = \|X\|^2$ in the heteroscedastic scenario.
This error distribution allows a non-trivial effect of higher moments 
because the (conditional) skewness  of the error $\e$, given by $\mathsf{skew}[\e|X] = (e^{\tau^2} + 2) (e^{\tau^2}-1)^{1/2}$, exponentially increases over different parameters $\tau^2$, 
while the lower moments $\eo[\e|X] = 0$ and $\eo[\e^2|X] = \s^2(X)$ do not depend on $\tau^2$. We  report results  with $\tau^2 = 0.1$ and $\tau^2=3$, which respectively give   skewness 1 and 100, approximately.

Based on 1000 Monte Carlo iterations per sample size and model configuration,
we compute the empirical coverage rates, as well as average widths, of   95\% confidence intervals for a projection $\langle \beta, X_0 \rangle$ from the symmetrized intervals from each bootstrap methods \cite[cf.][]{Hall88}; for example,  we construct WB intervals from approximating the absolute value of projection statistic in \eqref{WBeq_Tn}  through the absolute value of bootstrap projection statistic in \eqref{WBeq_TnStar}.
The three types of WB multipliers listed in \autoref{egW} are  considered and denoted as   WB1 (two-point), WB2 (standard normal), and WB3 \citep{mam93}, respectively.
We compare these WB intervals 
against   RB and PB intervals as well as intervals by a direct standard normal (e.g., CLT) approximation (cf.~Section~\ref{WBssec_2_2}).
Bootstrap methods are implemented based on 1000 resamples and truncation levels $k_n = g_n= [1.5 \cdot n^{1/6}] \in \{3,4,6\}$,   $h_n \in \{1, \dots, 15\}$,
where $[\cdot]$ denotes the closest integer, where  values of $k_n$ and $g_n$ are motivated by consistency conditions for  $\hat{\beta}_{k_n}$.

\subsection{Performance of WB bootstrap intervals} \label{WBssec_3_2}

\begin{figure}[!b]
	\centering
	\includegraphics[width=0.99\linewidth]{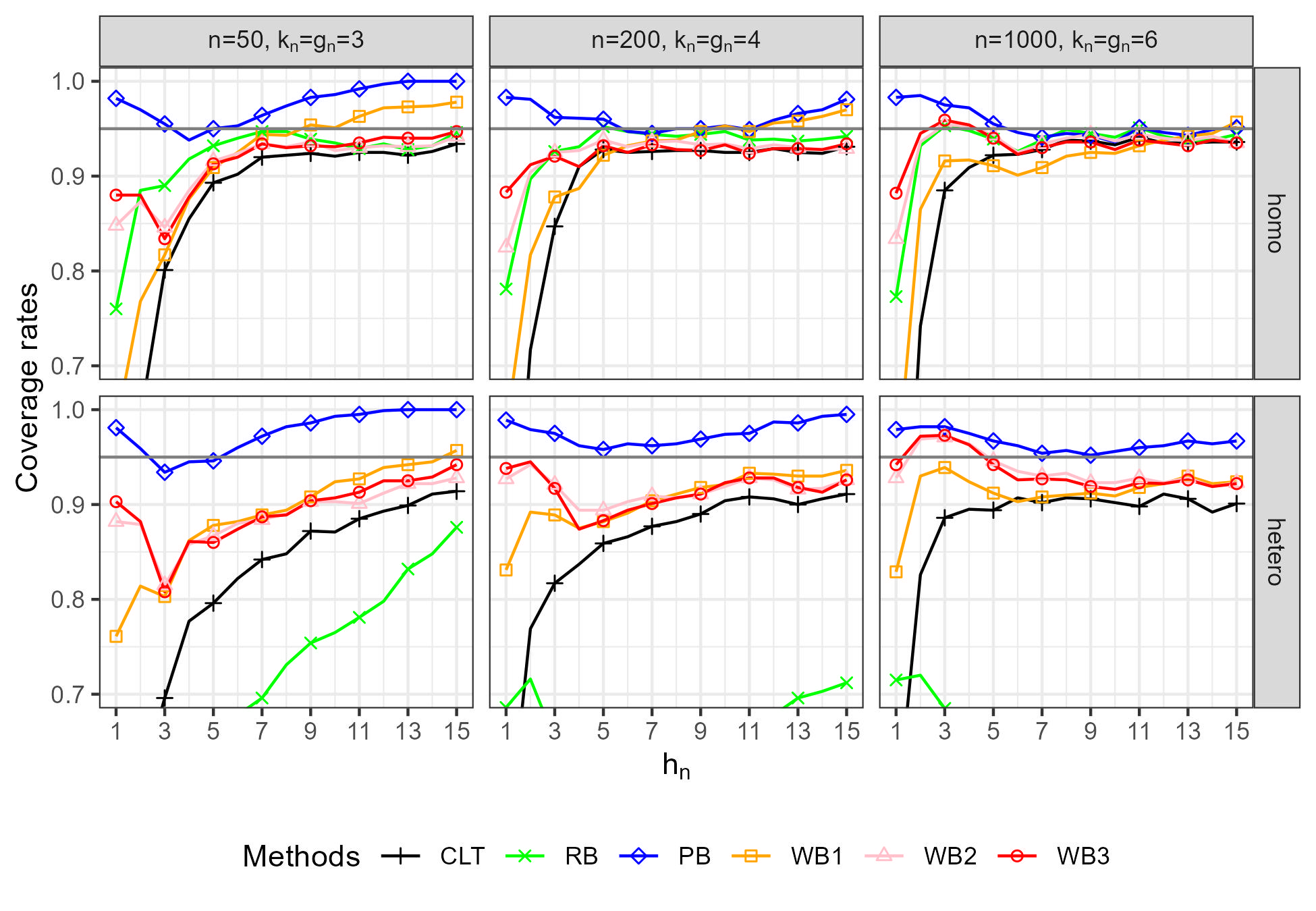}
	\caption{
		Under log-variance parameter $\tau^2=1$, empirical coverage rates of   95\%  intervals for the projection $\langle \beta, X_0 \rangle$  
		over   truncation levels $h_n \in \{1, \dots, 15\}$.
	}
	\label{fig1cover_dep_type}
\end{figure}

\begin{figure}[!b]
	\centering
	\includegraphics[width=0.99\linewidth]{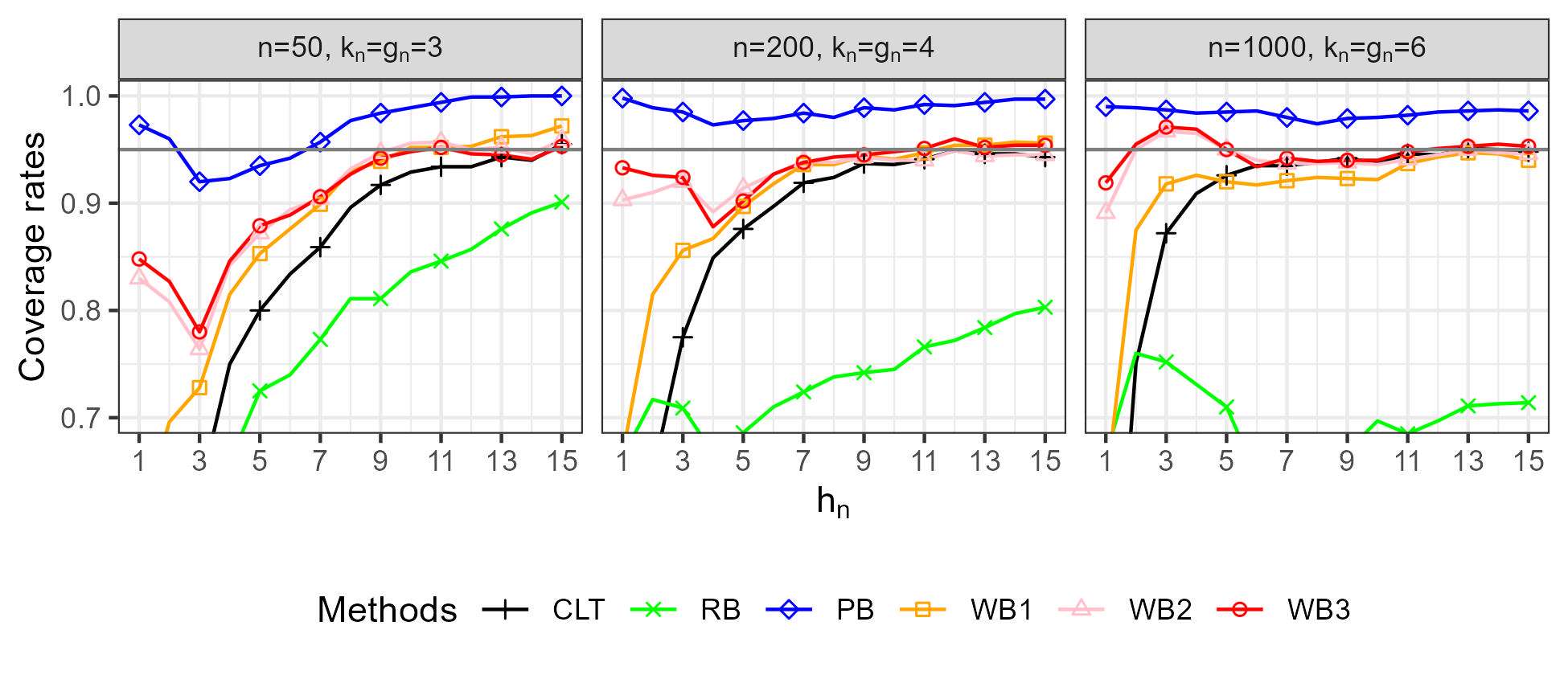}
	\caption{
		Under log-variance parameter $\tau^2=2$, empirical coverage rates of   95\%  intervals for the projection $\langle \beta, X_0 \rangle$  
		over   truncation levels $h_n \in \{1, \dots, 15\}$.
	}
	\label{fig2cover_skew_type}
\end{figure}

\autoref{fig1cover_dep_type}  shows 
empirical coverage probabilities over both scenarios of homoscedastic and heteroscedastic errors, focusing 
on a log-variance parameter  $\tau^2 = 0.1$  (or error skewness of 1).
Under homoscedasticity, WB intervals exhibit  accuracy that closely mirrors that of the RB intervals.
Under heteroscedasticity,   WB intervals  maintain relatively good  performance in coverage, while RB intervals are relatively poor in contrast (i.e., coverages for RB are so low as to often not plot in Figures~\ref{fig1cover_dep_type}-\ref{fig2cover_skew_type}) and   CLT-based approximation intervals also perform poorly.    PB intervals
tend to have better accuracy than WB intervals under heteroscedasticity    over smaller sample sizes $n  \in \{50,200\}$, though   
WB intervals  achieve comparable coverage accuracy for increasing sample sizes.

\begin{figure}[!b]
	\centering
	\includegraphics[width=0.99\linewidth]{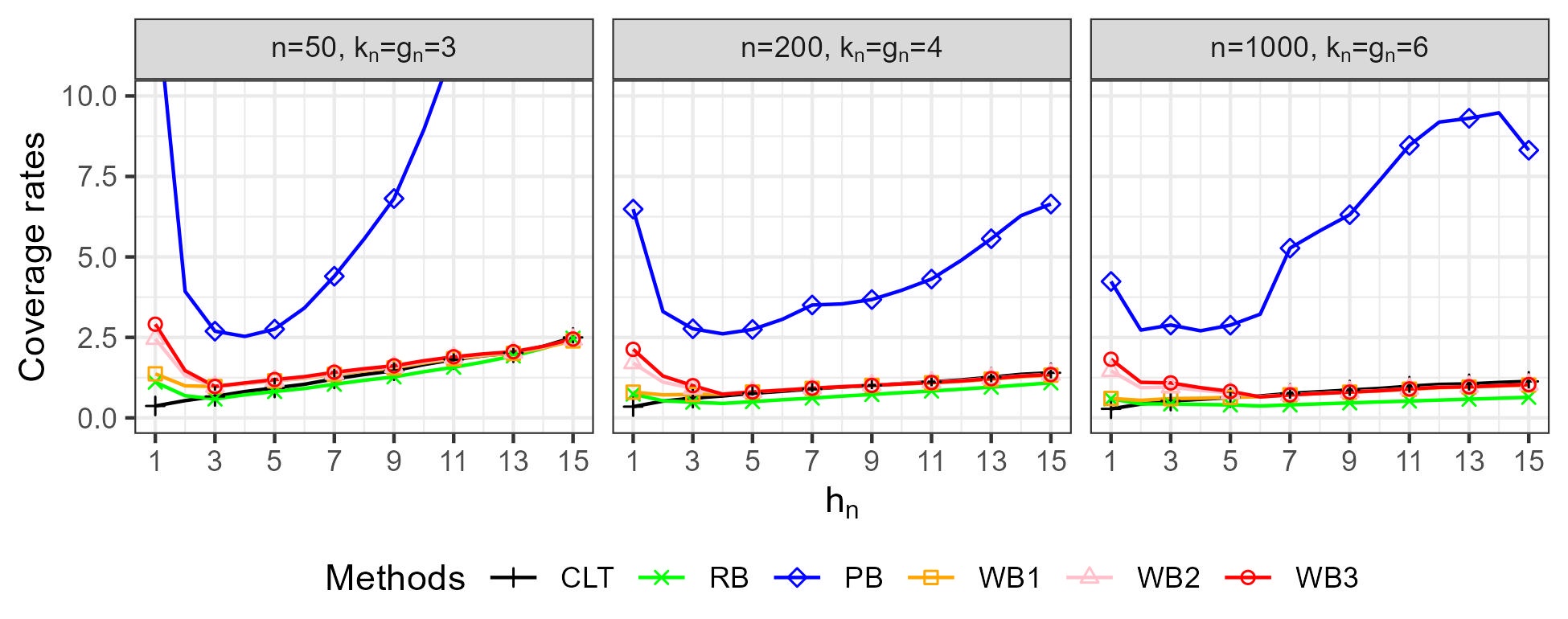}
	\caption{
		Under log-variance parameter $\tau^2=2$, average lengths of   95\%  intervals for the projection $\langle \beta, X_0 \rangle$  
		over   truncation levels $h_n \in \{1, \dots, 15\}$. 
	}
	\label{SfigSwidth_skew_type}
\end{figure}

Figure~\ref{fig2cover_skew_type} summarizes empirical coverages, focusing on heteroscedastic errors with a large log-variance parameter  $\tau^2 = 3$ (or error skewness of 100).   In contrast to the lower skewness case in Figure~\ref{fig1cover_dep_type},  the coverage results in Figure~\ref{fig2cover_skew_type} indicate that the WB can be preferable to PB in terms of coverage accuracy under heavy skewness.  In this setting, PB intervals can become rather conservative, as indicated in 
Figure~\ref{SfigSwidth_skew_type}
which shows average lengths of intervals in connection to  Figure~\ref{fig2cover_skew_type}. Here, the lengths of PB intervals are longer and less stable  
compared to WB intervals, which may owe to the fact that the PB method resamples regressor variables, unlike WB, which may create  more difficulties under heavy skewness.     
%Hence, a large level of skewness in errors can become a motivation for WB. 

\clearpage

	As raised by a reviewer,  regarding the effect of the truncation level $h_n$, 
	WB does often perform   worse than  PB
	in terms of coverage accuracy
	for small values of $h_n$ in Figures~\ref{fig1cover_dep_type}-\ref{fig2cover_skew_type}.   This aspect  attributes to  non-parametric resampling scheme of the PB, which can induce wider (more conservative) 
	intervals for such $h_n$ values (cf.~Figure~\ref{SfigSwidth_skew_type}).   However, these   results   also generally suggest that   
	WB intervals need higher truncation levels $h_n$ to become stable in length, after which these intervals also become more stable in coverage.     
	This insight can be helpful toward selecting a truncation level $h_n$ for the WB in application;  we revisit this point in \autoref{WBssec_4_1}  
	where practical implementation is discussed.
	In contrast, the length behaviors in Figure~\ref{SfigSwidth_skew_type}
	(as well as other length studies 
	in the supplement, cf.~Figure~S1) % \ref{SfigSwidth_dep_type}) 
	indicate that PB intervals tend to be 
	more sensitive than WB to the choice of the truncation level $h_n$, where   instability is   apparent when sample sizes are small  and the error skewness is strong.   The tuning parameter $h_n$ for PB remains  challenging to choose, and intervals can be quite wide.

\begin{rem} \label{rem_sim_kg}

		Following a  reviewer's suggestion,
		we conducted additional numerical studies 
		to examine the effects of $k_n$ and $g_n$ in greater detail.
		Specifically,   simulations   were conducted under the same setup as in \autoref{fig2cover_skew_type},
		but by fixing either $k_n$ and $g_n$ while varying the other as well as $h_n$. 
		The results are represented in Figure~S5
%		\autoref{figCoverVaryTune} 
		of the supplement. 
		Overall, large differences between $g_n$ and $k_n$ tend to produce over- or under-coverage in WB intervals,
		while $g_n$ close to $k_n$ leads to more reasonable behavior in  WB intervals.
		This finding supports  our WB implementation with $g_n$ being equal to $k_n$   in \autoref{WBssec_4_1}.
		We refer to Section~S3.3
%		\autoref{WB_ssec_S_kg}
		of the supplement for more details.
	
\end{rem}

\begin{rem} \label{rem_nonstd}

		One could construct WB statistics akin to \eqref{WBeq_TnStar} but without bootstrap-level studentization, 
		i.e., applying original data scaling $\hat{s}_{h_n}(X_0)$ from \eqref{WBscalingHeteroHat} in  place of bootstrap scaling $\hat{s}_{h_n}^*(X_0)$ from \eqref{WBeq_WBsHatStar}.  
		Additional numerical studies have indicated  that the coverage accuracy of this WB version is not robust to the choice of the multipliers, potentially because this WB does not mimic the actual studentization in the target statistic  (\ref{WBeq_Tn}).
		We refer to Section~S3.2
%		\autoref{WBssec_S4_2} 
		of the supplement for more discussion.
	
\end{rem}

\begin{figure}[!b]
	\centering
	\includegraphics[width=0.99\linewidth]{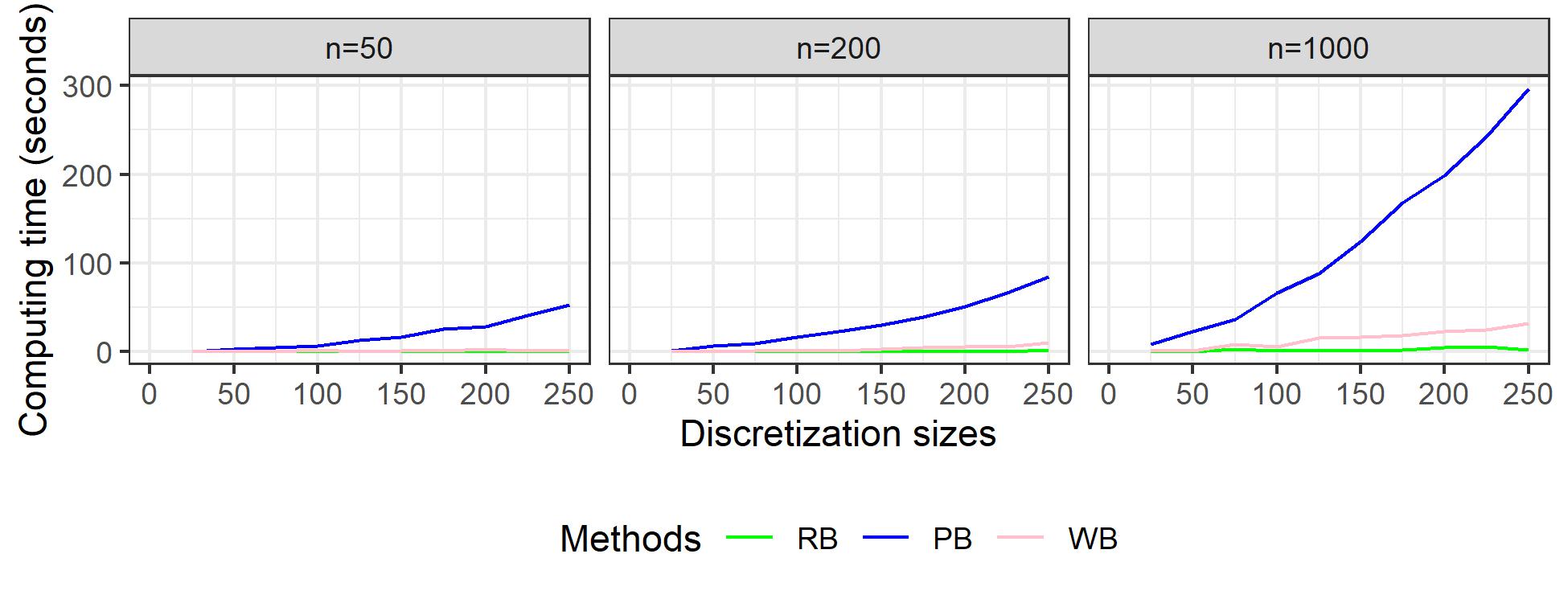}
	\caption{
		Computing timings of each bootstrap method in seconds over discretization sizes $M \in \{25, 50, \dots, 225, 250\}$.
	}
	\label{fig3timing}
\end{figure}

To  examine computational speeds among bootstraps for functional regression,  we consider the effects of different sample sizes $n \in \{50, 200,  1000\}$ and ten different discretization sizes $M \in \{25, 50, \dots, 225, 250\}$ for regressor curves.   Bootstrap confidence intervals were computed using three functions \texttt{RBinFLRM}, \texttt{PBinFLRM}, \texttt{WBinFLRM} found in the R package \texttt{BTSinFLRM}, as companion software to this paper, in  one representative simulation scenario 
($\s^2(X) = \mathrm{tr}(\ga)$ and $\tau^2 = 1$ with truncation levels $k_n = g_n=[ 2n^{1/6} ]$   and $h_n=g_n+1$).
Computing speeds for bootstrap intervals were timed 
on a machine equipped with Intel Xeon Gold 6144 Processor (with base frequency 3.5GHz)
using the R package \texttt{microbenchmark} \citep{microbenchmark}.
\autoref{fig3timing}  displays the average timing speeds from 1000 simulations. 
As expected,  WB and RB require similar amounts of computing time, due to their   resampling constructions with fixed regressors. 
However,  PB can be more time demanding than other bootstrap methods, particularly when either sample or discretization sizes is large,  owing to its more involved resampling scheme   with regressors. 
In this sense, WB can   have computational advantages over PB in heteroscedastic cases  with  large samples or fine discretizations.  
%Lastly, Figures~\ref{SfigStdEffectCover}-\ref{SfigStdEffectWidth} in the supplement \cite{supp} show that the studentization in WB can lessen the effects of the multipliers while the studentized PB intervals tend to be wider as the error distribution becomes more skewed. 
%A detailed discussion about studentization can be found in the supplement \cite[\autoref{WBssec_S4_2}]{supp}.

\subsection{Summary} \label{WBssec_3_3}

%While our proposed WB can often be comparable to existing bootstrap methods and  even superior   under certain settings,
%our goal is not to determine a universally superior bootstrap method but rather to recognize the relative merits of each approach and to frame the contribution of WB.
%Based on the numerical results, 
%we may provide general recommendations for choosing between the three bootstrap (PB/RB/WB) methods in functional linear regression. 
%Firstly, RB is  generally preferred when there is evidence of homoscedasticity,
%as this method often produces the best and most stable performances in coverage accuracy in this case, though the performance of WB can be comparable.   Under potential heteroscedasticity,  RB 
%should not be considered. 
%As PB and WB seem to work   well for most cases of both homoscedasticity and heteroscedasticity,
%these bootstraps can   be  good default choices.
%In particular, the small sample size performance of PB often turns out to be good, 
%often maintaining nominal coverage but 
%with coverage errors tending toward conservatism (over-coverage),
%while the other bootstrap methods can more generally exhibit under-coverage.  
%However, as seen in timing experiments, the PB  method can be computationally expensive for large sample sizes or fine discretizations of regressors.
%In such cases, WB is a natural alternative choice, as this approach is computationally fast and gives relatively good coverages for sufficiently large $h_n$, even when heavy skewness exists in errors.

	We emphasize that the proposed WB is designed for mean response inference without the intent to outperform the existing RB and PB. 
	Rather, the WB serves as an useful alternative to both RB and PB by offering key benefits: 
	(i) WB applies to both homoscedastic and heteroscedastic models, a clear benefit relative to the RB; 
	(ii) for large datasets, it shows both methodological and computational advantages over the PB; 
	(iii) WB performs better than the PB for some complex cases, e.g., for largely skewed errors, 
	where PB intervals are excessively conservative; and (iv) 
	WB shows  stability in interval length as a function of tuning parameters, much more so than PB, which helps toward selection of these parameters in practice (cf.~\autoref{WBssec_4_1}).

% ================= Section 4 =================================
\section{Practical implementation} \label{WBsec4}

\autoref{WBssec_4_1} overviews   choices of the truncation levels $k_n,g_n,h_n$ in practice with WB inference for mean responses.
This approach is then illustrated in \autoref{WBssec_4_2} through a United States weather dataset.

\subsection{Selection of truncation levels}  \label{WBssec_4_1}

Tuning parameter selection is an important problem in  regularization problems, including the methods based on the FPCR estimator. 
There have been some attempts to define an optimal truncation level based  on estimation error  \citep{HH07, CH06, CZZ18},
though such choices have no optimality guarantees for bootstrap inference.   
Additionally, in our context with bootstrap, appropriate truncation levels may also depend on the mean response (or projection) target of interest at a regressor $X_0$ value.
An optimal choice for bootstrap coverage accuracy requires more investigation, outside the scope of the current study. 
%Hence, though valuable, a fully theoretical investigation of choosing the truncation levels for our WB procedure should warrant a separate paper, like prior works \citep{Lahiri99, NL04, Nord09} in other resampling contexts, and we leave this direction for future exploration.

Instead, we suggest some practical guidance for choosing truncation levels $k_n,h_n,g_n$ with the proposed WB procedure based on theoretical and numerical findings. 
Firstly, the truncation $k_n$  for calculating residuals may be chosen by cross-validation (e.g., based on prediction error), 
as the requirement for the corresponding estimator $\hat{\beta}_{k_n}$ is simply consistency (cf.~\autoref{WBthmWB}).
%	 and \citet[Theorem~S3]{YDN24PBsupp}).
Secondly, we set $g_n=k_n$,
because $\hat{\beta}_{g_n}$ also requires only consistency
and the WB intervals perform well for $g_n$ equal to $k_n$ in our numerical studies
(cf.~\autoref{rem_sim_kg}).
Lastly, for $h_n$ used to infer the mean response $\mu(X_0)$,
we recommend selecting a value larger than $g_n$ for the following reasons:
the theoretical result in \autoref{WBthmWB} advises against choosing $h_n$ smaller than $g_n$ (i.e., by assumption (R) in the supplement), while
numerical  findings in \autoref{WBsec3} indicate that the WB intervals   perform better  and also exhibit stability for sufficiently large $h_n$ exceeding $g_n$. 
The latter point suggests that a value of $h_n$ may be selected by inspecting where WB intervals appear stable.   
With this in mind, we suggest a   concrete rule below for selecting $h_n$.
The strategy follows the spirit of the so-called  ``minimum volatility method"  of  \citet[Section~9.3.2]{PRW99}, 
which   aims to compute intervals over a series of bandwidths and then select the one where the resulting interval appears to exhibit the lowest volatility. 
Our heuristic also closely resembles the notion of    choosing  a truncation parameter where the fraction of variance explained in classical FPCA  appears stable   \cite[e.g.,][Section~12.2]{KR17}.

To elaborate further, with $k_n=g_n$ chosen as described above, 
let $H$ denote a finite set of consecutive candidate integers for selecting $h_n$ (e.g., $H = g_n+\{0,1, \dots, 20\}$).
We write $w_h$ and $c_h$ for the width and center of the WB intervals corresponding to $h \in H$.
To identify truncation levels where these two interval characteristics stabilize (i.e., widths and centers cease to change dramatically),
we consider the following sets:
\begin{align*}
	H_{w} 
	\equiv \{h \in H : |w_{h+1}-w_h|\leq \rho_w\},  \qquad 
	H_{c} 
	\equiv \{h \in H :|c_{h+1}-c_h|\leq \rho_c\},  
\end{align*}
where $\rho_w,\rho_c >0$ are some small thresholds for defining stability.
Denoting $H_{\mathrm{stable}} \equiv H_{ w} \cap H_{ c}$ and picking a small integer $r\geq 0$,
we determine a truncation level $h_n$ as
\begin{align}
	\hat{h}_n = \hat{h}_n(r) \equiv \min\{h \in H_{\mathrm{stable}}: h, h+1, \dots, h+r \in H_{\mathrm{stable}}\}.  \label{WBeq_truncOpt}
\end{align}
	Our selection strategy is  called the stabilized volatility method (SVM) hereafter.
	
\begin{figure}[b!]
	\centering
	\includegraphics[width=0.9\linewidth]{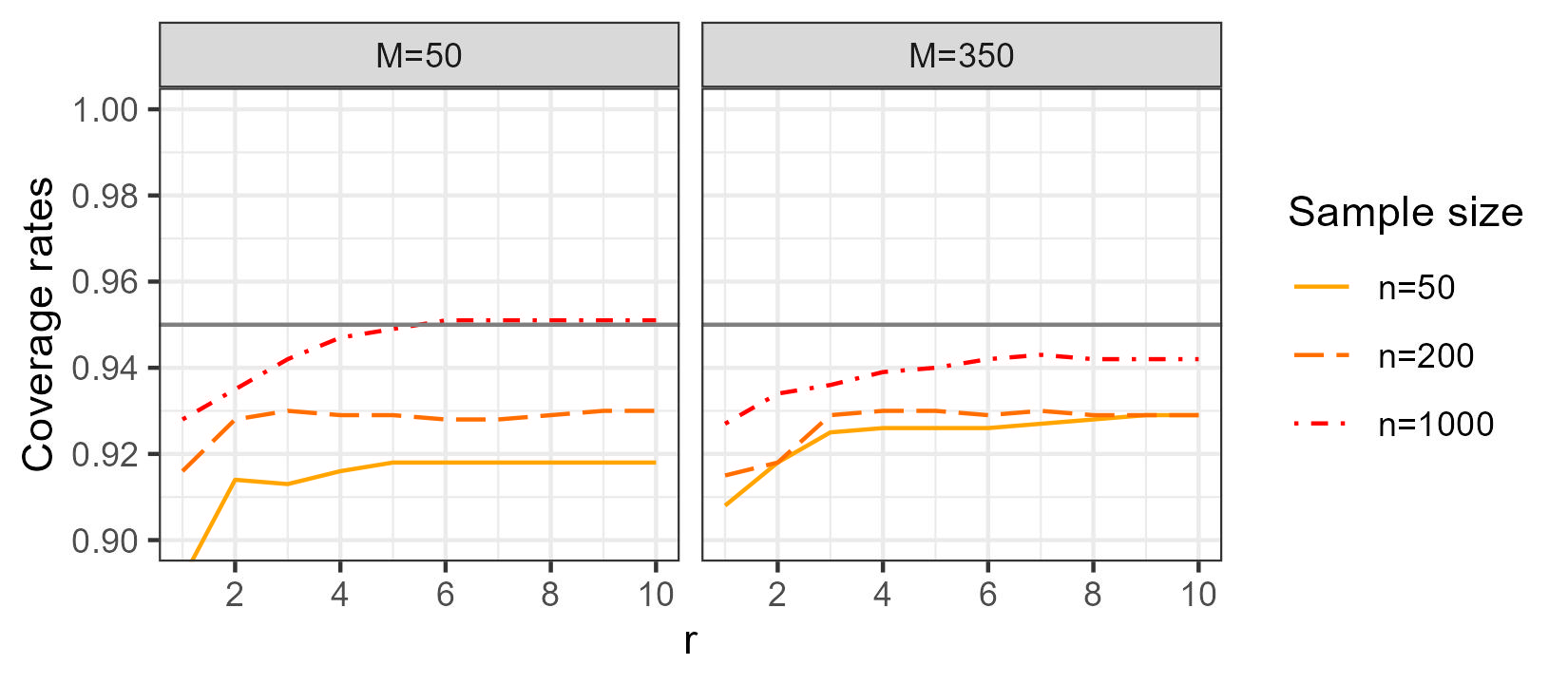}
	\caption{
		Empirical coverages of 95\% WB   intervals based on standard normal multipliers and   truncation $\hat{h}_n(r)$  selected by SVM   (\ref{WBeq_truncOpt}).
		Data generation matches that of \autoref{fig2cover_skew_type},
		where scenarios with a denser  regressor grid are also considered (right).
	}
	\label{figCoverSVM}
\end{figure}

	Using the same simulation setup as that for \autoref{fig2cover_skew_type},
	we conduct numerical studies to demonstrate the coverage performance of WB intervals based on SVM-selected truncation levels.
	With $\rho_w = \rho_c = 0.01$,  \autoref{figCoverSVM} displays empirical coverage rates of WB intervals over different $r \in \{1,2,\dots,10\}$ for defining the criterion  (\ref{WBeq_truncOpt}); 
	here we consider three different sample sizes $n$
	along with two discretization levels $M \in\{50,350\}$ for regressor curves.
	Note that the coverage results appear reasonable,   showing a stable pattern of behavior  over various $r$ for all sample sizes $n$ or levels $M$.  These findings suggest that the proposed SVM  
	for selecting the truncation level $h_n$ with WB can be useful. 
	In the next section,
	we further demonstrate this selection approach, which can be combined with a graphical visualization.

\subsection{Real data analysis} \label{WBssec_4_2}

To illustrate the proposed WB method,
we  consider a US weather dataset, called nClimGrid-Daily, provided by the National Centers for Environmental Information.
The data  contain  daily temperatures (\textdegree C) and precipitation (mm) on gridded fields in the US.\footnote{\href{https://www.ncei.noaa.gov/products/land-based-station/nclimgrid-daily}{https://www.ncei.noaa.gov/products/land-based-station/nclimgrid-daily}}
In particular, we extract   data for 2022, where 
each functional regressor $X_i$ represents daily average temperatures of a state $i$ in the US,
while the associated response $Y_i$ is the total precipitation for $i=1,\ldots,47$; three states (Alaska, Alabama, and Hawaii) do not appear in the data so that the total sample size is $n=47$. 
As  daily observations are available, the discretization size is large as $M=365$. 
We  examine four new regressor curves $\{X_{0,l}\}_{l=1}^4$ as the average curves of a region in the US: Northeast, South, North Central, and West as defined by the United States Census Bureau.  The supplement shows the temperature curves of each state and their averages $\{X_{0,l}\}_{l=1}^4$.

\begin{figure}[b!]
	\centering
	\includegraphics[width=0.7\linewidth]{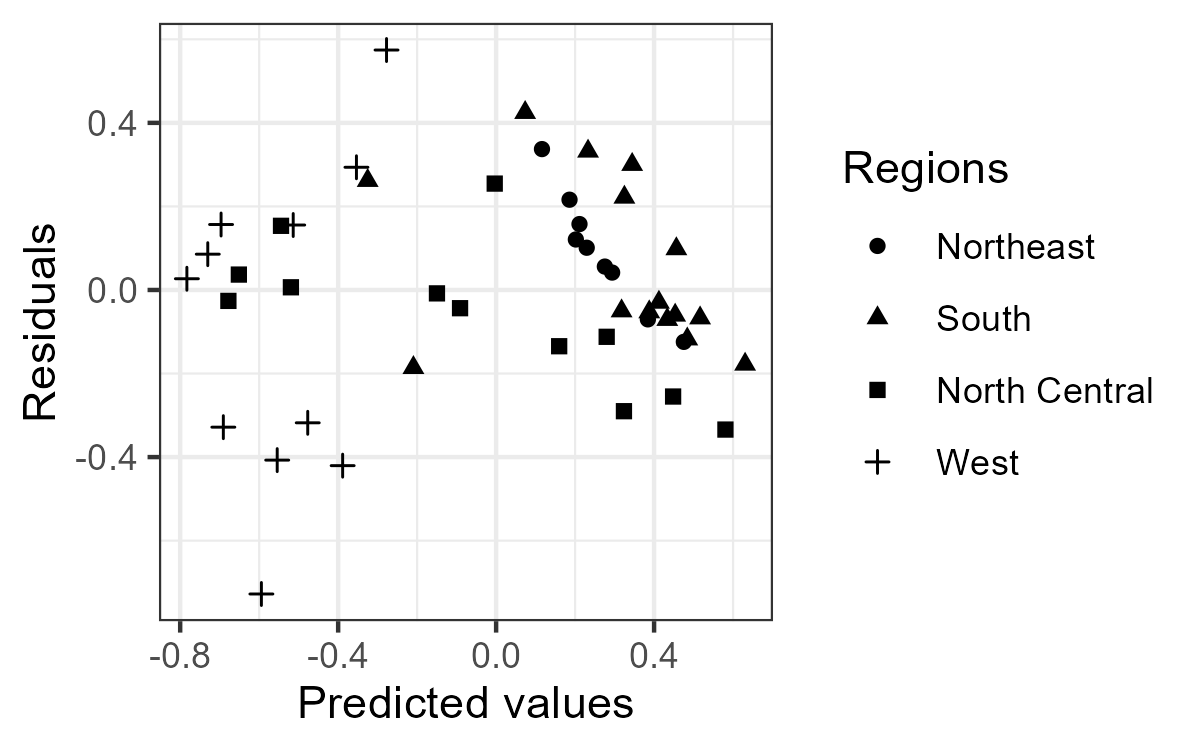}
	\caption{
		A scatter plot of residuals versus predicted values with $k_n = 6$ from the US weather dataset.
	}
	\label{figUSweatherResid}
\end{figure}

We first examine some distributional features of the data, in terms of heteroscedasticity and skewness, that may suggest the choice of  bootstrap method. A pilot truncation $k_n$ is first obtained as $k_n = 6$ by  cross-validation based on the prediction error as suggested in \autoref{WBssec_4_1}.
To examine whether the errors have constant variance over different regions,
we compute the residuals $\hat{\e}_{i,k_n} \equiv Y_i-\bar{Y} - \langle \hat{\beta}_{k_n}, X_i-\bar{X} \rangle$,
and \autoref{figUSweatherResid} provides a scatterplot of residuals versus the predicted values $\hat{Y}_{i,k_n} \equiv \bar{Y} + \langle \hat{\beta}_{k_n}, X_i-\bar{X} \rangle$.
This figure suggests some heterosceadsticity in the data, especially due to West region (as does a study of estimated standard deviations shown in the supplment).  
In the presence of potential heteroscedasticity, 
we might expect that RB intervals should not be used (or may work poorly) and that PB or WB intervals would be  more reasonable.  
We next consider kernel-estimated densities from  residuals in each region
to examine skewness; these (as shown in the supplement)   exhibit some slight skewness, again particularly in West region.
As discussed in \autoref{WBsec3}, potential skewness in errors can suggest that the WB method can become  
preferable to  PB for coverage accuracy. 
Further, as the discretization size $M=365$ is relatively large for these data, WB is also computationally faster to implement than PB, as shown in \autoref{fig3timing}.

We will apply the results in \autoref{WBthmWB} 
to construct (symmetrized) WB intervals of the centered projections $\{\langle \beta, X_{0,l}-\eo[X] \rangle\}_{l=1}^4$ for all four regions with normal multipliers.
This allows us to compare the original projections $\{\langle \beta, X_{0,l} \rangle\}_{l=1}^4$ with the global average projection $\langle \beta, \eo[X] \rangle$,
and hence, the mean responses $\{\mu(X_{0,l})\}_{l=1}^4$ with the global mean $\mu(X)$. 
That is, the resulting intervals have the interpretation of indicating how the mean response (total precipitation) per region may differ from a common global average.
For bootstrap inference, 1000 bootstrap resamples are used along with $g_n = k_n = 6$ as recommended in \autoref{WBssec_4_1}. 

\begin{figure}[b!]
	\centering
	\includegraphics[width=0.8\linewidth]{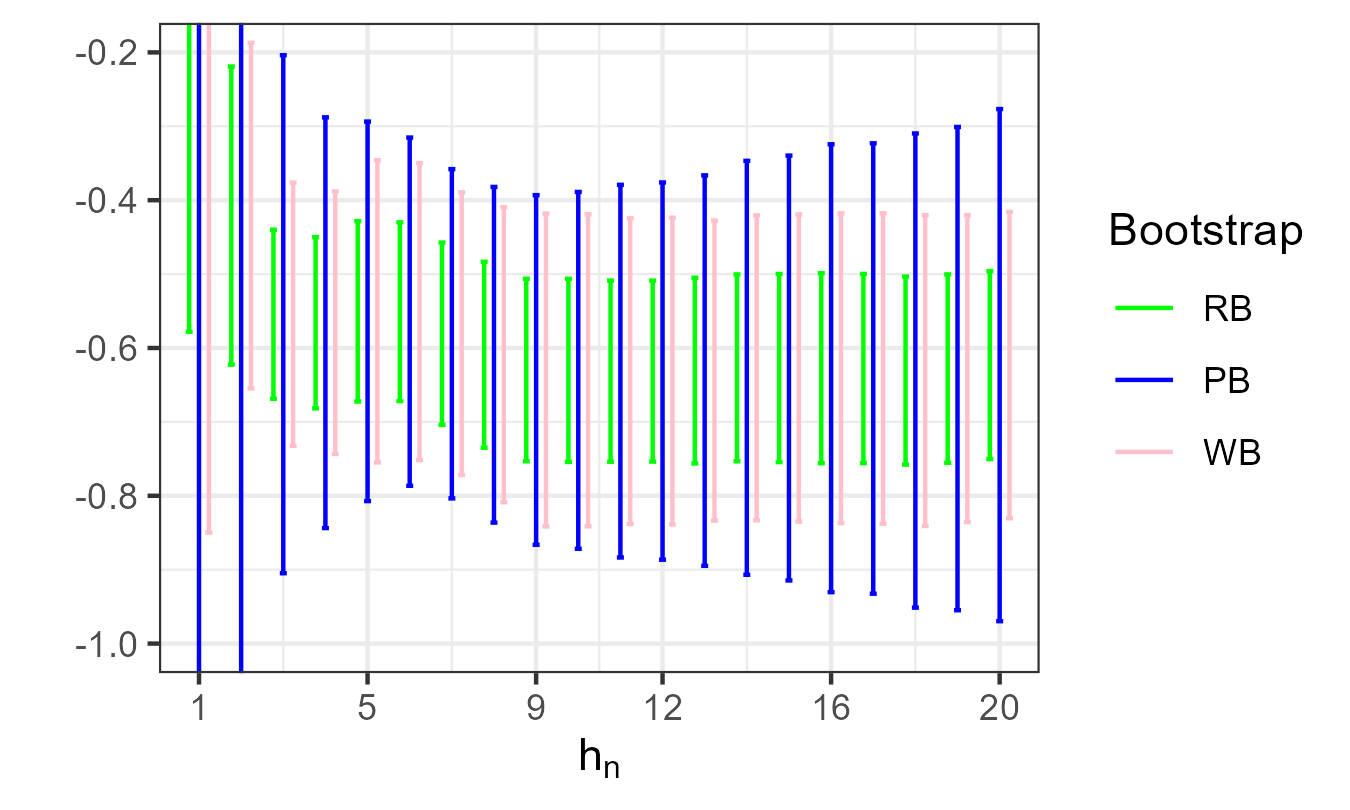}
	\caption{
		Bootstrap intervals for centered projection $\langle \beta, X_{0,4} - \eo[X] \rangle$ for West from the US weather dataset with $k_n = 6 = g_n$ and different $h_n \in H \equiv \{1, \dots, 20\}$.
		The plot is cropped within $[-1.0, -0.2]$ due to too large widths of PB intervals. 
	}
	\label{figUSweatherIntervalsWest}
\end{figure}

To illustrate  application of the selection rule for   $h_n$ from \autoref{WBssec_4_1} for WB intervals,
we focus on  the West region (i.e.,   the value of $X_{0,l}$ for $l=4$) to guide a choice of $h_n$, where heteroscedasticity and skewness are more pronounced.     
The bootstrap intervals for the West region, computed for different values of $h_n$, are presented in \autoref{figUSweatherIntervalsWest}.
RB intervals are  narrow in general,
	though less accuracy  for this method would be expected under heteroscedasticity.
	The widths of the PB intervals tend to be wide and exhibit the least stable pattern in widths;
	this also matches behavior observed in numerical studies (Section~\ref{WBsec3}).
	Note that the WB intervals stably behave like the RB counterparts, though with larger widths to incorporate heteroscedasticity. 
%As expected, the WB intervals are narrower than the PB intervals but wider than the RB intervals. 
\autoref{figUSweatherIntervalsWest} visually suggests that the WB intervals (pink) begin to stabilize in both center and width, 
when the truncation level $h_n$ reaches 9 or higher.

To 
quantitatively select $h_n$ by the SVM  of \autoref{WBssec_4_1}, 
we compute $\hat{h}_n = \hat{h}_n(r)$ in \eqref{WBeq_truncOpt} with $H \equiv \{6,\ldots,20\}$ and  $\rho_w = \rho_c = 0.01$ for the candidate sets.
	It is worthwhile recalling from the right panel of \autoref{figCoverSVM} that
	the proposed SVM was demonstrated as reasonable,
	even for densely observed regressors as in  these weather data.
Our findings indicate that, for any $r \in \{1, \dots,10\}$, the selected truncation level for the West region is $\hat{h}_n = 9$,
as also visually identified from \autoref{figUSweatherIntervalsWest}.
Figure~S9 of the supplement shows the intervals for the other regions based on each bootstrap,
where the best empirical choice of $h_n$ for the inference of $\mu(X_{0,l})$ may  change with or depend on the new regressor $X_{0,l}$ (i.e., region)
and the selected values  for WB were 9 for Northeast, 12 for South, and 10 for North Central, e.g., for all $r \in \{2,3,4,5,6\}$.

In \autoref{tbUSweatherIntervalOpt}, we provide the WB intervals based on the chosen truncation levels $h_n$.
The results support a conclusion that the projections $\langle \beta, X_{0,l} \rangle$ from each region differ substantially from the global mean projection $\langle \beta, \eo[X] \rangle$, 
as none of the WB intervals contain zero.
More specifically, the mean total precipitation in the Northeast and South appears higher than the global mean,
whereas the North Central and West regions experience less rainfall on average compared to the the total precipitation across all 47 observed states. See the supplement for further numerical summaries.   As these data   seemingly involve complicated error structures (i.e., with  heteroscedastic and skewed distributions), 
the WB intervals   are helpful toward drawing valid inferences. 

\begin{table}[h!]
	\centering
	\caption{
		95\% WB intervals for centered projections $\{ \langle \beta, X_{0,l} - \eo[X] \rangle \}_{l=1}^4$ from  weather data using selected truncations $k_n = 6 = g_n$ and $h_n=9$ for Northeast ($l=1$), $h_n=12$ for South ($l=2$), $h_n=10$ for North Central ($l=3$), and $h_n=9$ for West ($l=4$).
	}
	\label{tbUSweatherIntervalOpt}
	\medskip
	\setstretch{1.2}
	\begin{tabular}{cccc} \hline
		Northeast & South & North Central & West \\ \hline
		$[0.239, 0.466]$ & $[0.267, 0.445]$ & $[-0.220, -0.029]$ & $[-0.842, -0.418]$ \\ \hline
	\end{tabular}
\end{table}

% ================= Section 6 =================================
\section{Concluding remarks and outlook} \label{WBsec5}

Uncertainty quantification  is an important, but challenging, problem  in functional linear regression models (FLRMs). 
We   proposed and formally established a new   wild bootstrap (WB) method for mean responses in FLRMs.
As suggested by both theoretical findings and numerical results, 
WB combines the strengths of both residual bootstrap (RB) and paired bootstrap (PB) with functional regressors: 
WB retains a low computational burden by keeping regressors fixed in the bootstrap scheme, similar to RB, 
while retaining validity under both homoscedasticity and heteroscedasticity, akin to PB.
In particular, WB serves as a useful alternative to RB in heteroscedastic cases and to PB when the data structure is large, 
while also outperforming both methods in  coverage accuracy under heavy skewness in heteroscedastic errors and also allowing for a selection method for tuning parameters due to stability in WB interval behavior.
Additionally, we have prepared an accompanying R package, 
which provides implementations of WB as well as RB and PB for setting confidence intervals for mean responses in FLRMs.

	We conclude   by outlining several potential extensions of interest. 
	As  \cite{KH24} and \cite{LL25} apply bootstrap for testing the slope $H_0:\beta=0$ in homoscedastic FLRMs,  it may be possible to consider WB for extending such slope tests under heteroscedastic errors.     Further, 
	bootstrap developments could also be meaningfully considered for inference about  more complicated FLRMs such as  with functional response \citep{DT24} or as with generalized \citep{MS05}, partial \citep{Shin09}, or high-dimensional \citep{XY21} FLRMs.
	Finally, further investigation is warranted in developing bootstrap inference in FLRMs with sparsely observed functional regression due to unique challenges from sparseness \citep{ZYZ23}.
	Such   research   could enrich  inference in FLRMs.

%\newpage
\bibliographystyle{dcu}
\bibliography{BTSinFLRM3WB}

\end{document}